\documentclass{aa}
\usepackage{graphicx}
\usepackage{txfonts}
\begin{document}
   \title{The Birth-Cluster of the Galactic Luminous Blue Variable  WRA~751\thanks{Based on observations obtained with ESO telescopes on La Silla and Paranal, Chile, under programmes 271.D-5008, 71.C-0432, and 073.D-0037}}

   \author{A. Pasquali
          \inst{1}\inst{2}
          \and
          F. Comer\'on\inst{3}
	  \and
	  A. Nota\inst{4}}

   \offprints{A. Pasquali}

   \institute{Institute of Astronomy, ETH H\"onggerberg, HPF,
              CH-8093 Z\"urich, Switzerland\\
         \and
              Max-Planck-Institut f\"ur Astronomie, K\"onigstuhl 17,
              D-69117 Heidelberg, Germany\\
              \email{pasquali@mpia.de}
         \and
             ESO, Karl-Schwarzschild-Strasse 2, D-85748 
	     Garching bei M\"unchen, Germany\\
	     \email{fcomeron@eso.org}
         \and
	     STScI, 3700 San Martin Drive, MD28201 Baltimore, USA\\
	     \email{nota@stsci.edu}
             }

  \date{Accepted November 8, 2005}

   \abstract{We present the results of NTT/VLT $UBV$ imaging of a 260 arcmin$^2$
   region containing the Galactic Luminous Blue Variable WRA~751, in search for its 
   birth-cluster, i.e. a cluster of young and massive stars spatially and physically 
   associated with it. On the basis of the classical reddening-free parameter $Q$,
   we have identified a sample of 24 early-type stars with colours typical of
   spectral types earlier than B3. Interestingly, these stars are clustered within 
   a radius of 1$'$ from WRA~751, corresponding to about 1$\%$ of the imaged field.
   These stars tightly distribute around $(B-V) \simeq$ 1.67, which in turn defines a 
   mean extinction $A_V \simeq$ 6.1 mag. The 5 brighter ($V \leq$ 16.2) and bluer 
   ($Q \leq$ -0.9) stars of the sample have been subsequently observed with FORS1 and  
   classified as 3 late O- and 2 early B- stars. The absence of stars earlier
   than O8 indicates an age of the cluster older than 4 Myr, although it could be
   due to an incomplete sampling of the upper end of the main sequence.
   Nevertheless, the detection of OB stars of class I certainly indicates an age of
   a few million years. At an assumed distance of 6 kpc,
   we estimate a cluster radius of 3.4 pc and a total mass of 2.2 $\times$ 10$^3$ M$_{\odot}$.
   Our discovery is only the second known instance of a Galactic Luminous Blue Variable
   associated with its birth-cluster.
   }

   \maketitle
%

\section{Introduction}
Massive stars (M$_i \geq$ 40 - 60 M$_{\odot}$, i.e. progenitors 
of supernovae of type II) play a key role in galaxy
formation and evolution. Their stellar winds and end-of-life explosions
set the kinematics and chemistry of the interstellar medium,
and power galactic fontains and outflows, which, in turn, significantly
affect galaxy evolution (i.e. feedback, Dekel \& Silk, 1986).
This is true at any lookback time; however, beyond the Local Group,
individual stars are no longer resolved and one
only observes their effect integrated
over the entire galaxy. To interpret these observations,
population synthesis codes are required (i.e Bruzual \& Charlot 2003,
Leitherer et al. 1999). The accuracy of these models relies on our
detailed knowledge of the first step in the ladder, the massive stars 
evolution, and, in particular, the dependence of mass loss on initial 
stellar mass and time (cf. Langer et al. 1994). This is still under debate.

Theoretical predictions indicate that mass loss is a competition among
stellar gravity, radiation pressure and rotation (Maeder \& Meynet 2000).
Unfortunately, observations have failed so far to quantitatively constrain
the rotational velocity of evolved massive stars. Mass loss
becomes extreme in the post-main sequence evolution, when stars reduce
their mass by a factor of $\geq$ 4 and approach the final stage of supernova.
It is believed that massive stars remove the majority of their envelope
during the Luminous Blue Variable (LBV) phase, via giant eruptions that
produce circumstellar, chemically-enriched nebulae (i.e. Eta Carinae,
Humphreys \& Davidson 1994).

Nebular expansion velocities and nebular sizes set the lifetime of the
LBV phase to $\simeq$ 10$^4$ yrs (Nota 1997), and this is the
reason why only about 30 candidate and confirmed LBV stars are known in
the entire Local Group (Humphreys \& Davidson 1994, Nota 1997).
Of these, eleven belong to the Milky Way and only one confirmed LBV is 
located in a star cluster (Eta Carinae in Trumpler 14/16,
Massey et al. 2001), and a candidate LBV candidate, VI Cygn No. 12
is associated with Cgynus OB2. In the LMC, only one confirmed LBV 
out of 10 is a member of a cluster of massive stars (S Doradus
in LH~41, Massey et al. 2000). 

The presence of a cluster of OB stars associated with an LBV star gives us
the possibility to infer the distance, age and mass of the LBV progenitor by 
fitting the cluster Zero-Age Main sequence and its turn-off.  
In this way, Massey et al. (2001) have derived a progenitor mass of about 
120 M$_{\odot}$ and an age of about 1 Myr for both LBV stars. The authors have 
thus suggested that LBVs are a normal stage in the evolution of the most 
massive stars.

To further test this hypothesis better statistics are required, and the
discovery of the birth-cluster of a LBV star is thus extremely valuable. 
In this paper
we report on the discovery of a second case of association between a
confirmed galactic LBV and an OB cluster. The LBV star is WRA~751 in the 
Galactic Carina arm. Long-slit spectra of its circumstellar nebula
(Nota, private communication) revealed the presence of three emission-line (i.e. 
H$\alpha$) stars in the close vicinity of WRA~751. Unfortunately, the 
spectra only covered the wavelength 
range 6200 - 8000 \AA, and the 4000 - 5000 \AA\ interval, needed to 
derive reliable spectral types, was missing. We have 
therefore followed-up these early observations with UBV imaging of a
region centred on WRA~751 to identify other possible young, massive
stars on the basis of their $UBV$ photometry. We do confirm the existence of 
moderately reddened early-type stars and present spectra of a 
selection of them, which we classify as late O- and early B-type stars of 
various luminosity classes. The absence of stars earlier than O8 indicates
a cluster age older than 4 Myr.
\par\noindent
The observations and data reduction are presented in Sect. 2, while the identification
of the WRA~751 cluster is described in Sect.3. The photometric and 
spectroscopic
properties of the cluster stars are discussed in Sect.4, and conclusions follow in
Sect.5.


\section{Observations}

\subsection{Imaging}

  Our first set of imaging observations was obtained on the night of
1/2 April 2003 using EMMI, the visible imager and spectrograph at
the ESO New Technology Telescope on La Silla. The observations
were designed so as to include the region immediate to WRA~751 as
well as an extended, diffuse nebulosity visible in the UK Schmidt
Telescope H$\alpha$ survey (Parker 1997) extending to
its Southwest. An area of $18'.6 \times 11'.4$ centered on the
coordinates $\alpha(2000)=11^h 08^m 15^s5$, $\delta(2000) =
-60^\circ 47' 35''$ was covered in a mosaic of pointings through
the $U$, $B$, and $V$ filters. The $U$ and $B$ frames were
obtained on the EMMI blue arm, which uses a 
$1024 \times 1024$~pix$^2$ detector yielding a field of view of
$6'.2 \times 6'.2$ per frame with a scale of $0''.37$~pix$^{-1}$. The
$V$ frames were obtained with the EMMI red arm, which uses an
array of two $2048 \times 4096$~pix$^2$
detectors yielding a field of view of $9'.1 \times 9'.9$. We used $2
\times 2$ binning in the red arm, resulting in a scale of
$0''.333$~pix$^{-1}$. The field was covered in a sequence of ten
pointings in each filter consisting of telescope offsets of 3'
westwards, followed by offsets of 5' alternatively northwards and
southwards. The offsets ensured a broad overlap between adjacent
frames resulting in each sky position being recorded by at least
two ($U$ and $B$ filters) or three ($V$ filter) pixels. Exposure
times for each frame were 120~s ($U$), 45~s ($B$), and 10~s ($V$).
Additional observations were obtained of the
Landolt (1992) standard field around PG1047+003 in
order to determine the photometric zeropoints in each filter.
Characteristic extinction coefficients for La Silla were taken
from the observatory Web pages, and photometric system
transformation coefficients were obtained from the EMMI User
Manual.

The NTT observations were reduced using standard IRAF\footnote{IRAF
is distributed by NOAO, which is operated by the Association of
Universities for Research in Astronomy, Inc., under contract to
the National Science Foundation} tasks to carry out the usual
steps of bias subtraction, flat-field correction using twilight
sky flat field frames, removal of cosmic ray hits, and
construction of a bad pixel mask based on the pixel value
histograms of bias and flat field frames. Precise offsets among the
images composing the mosaics were determined using the position of
the numerous stars in the overlap areas between consecutive frames
as a reference. The images were then combined into
a single mosaic of the field by averaging the pixel values at each
sky position, discarding those flagged in the bad pixel mask.
Finally, the mosaics obtained in each filter were registered and
cropped to the common area.

  A second set of observations was obtained using FORS2,
one of the 
visible imagers and low-resolution spectrograph instruments at the ESO 
Very Large Telescope, in Service Mode on the night of 7/8 March 2004. The instrument
detector is composed by mosaic of two $4096 \times 2048$~pix$^2$ detectors 
yielding a pixel scale of $0''.253$~pix$^{-1}$ with the $2 \times 2$ binning 
used. A dither pattern focusing on the northern region of the WRA~751 
cluster, which was not included in our previous NTT observations, was 
observed through the $UBV$ filters. We used the pipeline-reduced individual 
images included by the Observatory in the data package (Silva \& 
P\'eron 2004) and stacked them into a single frame per filter 
for a given exposure time. Two exposure times per filter were used: 
5s and 120s ($U$), and 1s and 60s ($B$, $V$), so as to increase the 
dynamic range by obtaining 'shallow' and 'deep' images. The field 
covered by the FORS2 observations is $7'.06 \times 7'.06$ centered on 
$\alpha(2000) = 11^h 08^m 40^s4$, $\delta(2000) = -60^\circ 42' 51''$.
Photometric calibration was achieved by using calibration plan observations 
of a Landolt (1992) standard field in the SA 100 area containing 
star \#269, also reduced by the FORS2 pipeline and included in the 
data package. Extinction and colour coefficients were taken from the 
FORS2 Quality Control Web pages maintained by ESO 
({\tt http://www.eso.org/observing/dfo/quality/}).

  Instrumental photometry was carried out under IRAF by means of 
dedicated scripts that make use of tasks in the DAOPHOT package
(Stetson 1987). Sources were first detected with
DAOFIND, using first the images of isolated, unsaturated bright
stars to determine an approximate point-spread function (PSF),
needed for the identification of point sources. Photometry of all
the stars detected by DAOFIND in each frame was then performed by
defining an undersized aperture having as a radius the full-width
at half maximum of the PSF of the reference star, and measuring
the flux inside it at the position of each detected object using
the PHOT task under DAOPHOT. The rest of the flux in the PSF of
each star was then added by fitting a circularly symmetric radial
profile to the distribution of counts of each star outside the
aperture, rather than by using a constant aperture correction.
This procedure allowed us to remove the photometric contamination due
to other stars located on the wings of the PSF, and to
adjust to the mild variations of the image quality across the
combined images. For the FORS2 observations, separate catalogs 
of magnitudes for each filter were produced from the deep and shallow 
images, and the magnitudes were then compared to determine the onset of 
the non-linearity regime of the detector. The magnitude of each 
star was taken from the measurement in the deep image unless its 
magnitude was less than 0.5~mag fainter than the start of the 
non-linearity regime. For brightest stars the measurement on the shallow 
image was preferred instead. 

  A merged catalog containing the results from the EMMI and FORS2 
observations was finally prepared. The magnitudes measured with FORS2 
were preferred when available due to their higher quality, and the EMMI 
magnitudes were used in the area not covered by the VLT observations.

\subsection{Spectroscopy}
The spectroscopic observations were performed in Service Mode with the 
FORS1 instrument on the VLT using multi-object spectroscopy (MOS) on the 
night of 3/4 July 2003. A 
selection of targets was made from our EMMI
observations (which were the only ones available at the time) based on 
magnitudes brighter than $V \simeq$ 16.2 and colours bluer than $Q \simeq$ -0.9. 
The MOS masks in FORS1 are defined by a set of movable blades, which constrains 
the number of slits that can be placed on a small area of the sky such as 
that subtended by the WRA~751 cluster. With this setup we could obtain 
spectra of five OB star candidates (located within 1$'$ from WRA~751
itself) in a single 600s exposure. 
Spectra were taken through the GRIS$_{-}$600B$+$12 grism covering
the spectral range between 3450 \AA\ and 5900 \AA, where the
diagnostic H, He, C and N lines used in the MK spectral type 
classification scheme lie. The GRIS$_{-}$600B$+$12 grism is characterised by 
a dispersion of 1.2 \AA/pix and a FWHM resolution of 4.8 \AA.
\par\noindent
  The data were reduced with the standard routines 
in the IRAF APEXTRACT and ONEDSPEC packages.

\section{Identifying blue stars \label{ident_blue}}
At a galactic longitude $l \simeq 290^\circ .7$ our field probes
a line of sight providing a nearly end-on view of the
Sagittarius-Carina arm, which extends between $\sim 1.5$~kpc and
$\sim 8$~kpc in that direction (e.g. Georgelin \&
Georgelin 1976). At the low galactic latitude of the
field, $b \simeq -0^\circ .4$, the far end of the arm lies less
than 60~pc from the galactic plane. It is thus expected to find in
our field stars with highly varying degrees of extinction, leading
us to adopt reddening-free selection criteria to identify the
earliest-type stars in the region.

Figure 1 shows the $(B-V)$, $(U-B)$ diagram of all
the non-saturated stars in the field with colours determined to 
better than $\sigma (B-V)$, $\sigma(U-B) = 0.2$. The unreddened 
main sequence {\it locus} from Bessell (1990) is plotted, 
as well as a reddening vector showing the displacement of a B3V star 
obscured by $A_V = 5$~mag with a reddening law characterized by
$A_V = 3.1 E[B-V]$, $E[U-B] / E[B-V] = 0.80$. The adopted reddening 
ratio $E[U-B] / E[B-V]$, also preferred by Massey \& 
Thompson (1991) in their study of Cygnus~OB2, is somewhat 
larger than the standard 0.72. However, it is in better agreeement with 
the distribution of points in the $(B-V), (U-B)$ diagram, since a shallower 
reddening vector with a slope of 0.72 would leave many more stars in a 
region inaccessible by a reddened star with normal intrinsic colours.
We note in any case that the precise choice of the value of 
$E[U-B] / E[B-V]$ has virtually no impact on our results concerning the
detection of early-type stars. 
The total-to-selective extinction ratio that we have adopted, $R_V =
A_V / E[B-V] = 3.1$, is the classical one and very close to the one used
by Massey \& Thompson (1991), $R_V = 3.0$, for Cygnus OB2. The latter is
in turn a roundoff of the value obtained by Torres-Dodgen et al. (1991),
$R_V = 3.04 \pm 0.09$, which is virtually indistinguishable from the
classical value given the uncertainty.

\begin{figure}
\resizebox{8.5cm}{!}{\includegraphics{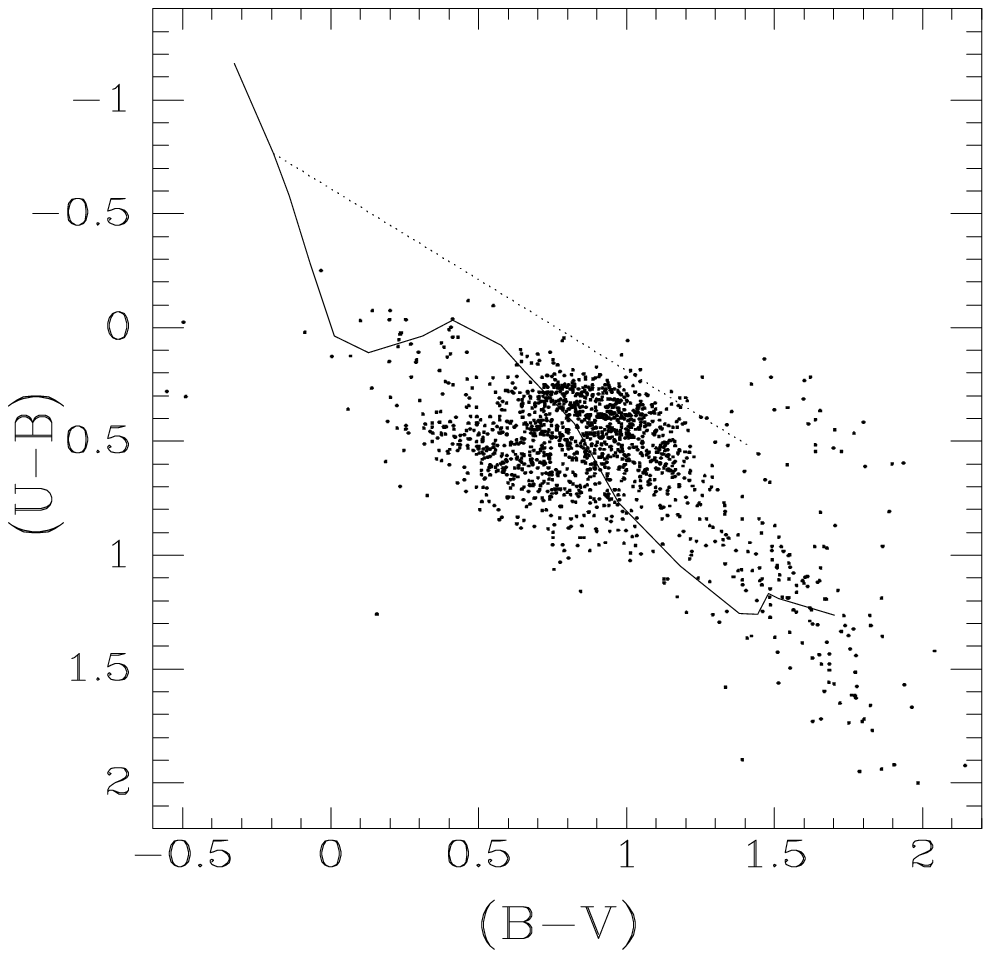}}
\par\noindent
{\bf Fig. 1.} $UBV$ colour-colour diagram showing the positions
of the non-saturated stars in our field having 
$\sigma (B-V)$, $\sigma(U-B) < 0.2$. The solid line is the {\it locus} 
of unreddened main sequence stars between spectral types O and M5 
(Bessell 1990). The dashed line is a reddening vector whose 
length corresponds to a visible extinction $A_V = 5$~mag.
\label{colcol}
\end{figure}

The bulk of stars in the field have positions in the colour-colour
diagram suggesting intrinsic colours in the range of the kink of
the solid curve (corresponding to spectral types between early A
and early G), reddened by $A_V \sim 2$~mag. Their magnitudes are
consistent with their location  at typical distances of $\sim
2$~kpc, in the near end of the Sagittarius-Carina arm. A tail of
redder stars, most probably cool luminous giants at varying
distances and reddenings, is also clearly visible. However, the
most remarkable feature related to our project is the loose clump
of stars roughly centered on $(B-V) \simeq 1.7$, $(U-B) \simeq
0.3$, a position that is accessible only to reddened stars of the
earliest types. Their location can also be characterized by the
classical reddening-free parameter $Q = (U-B) - 0.80 (B-V)$
(Johnson \& Morgan 1953)\footnote{Our definition
follows the usual convention in current literature, which reverses
the sign of the original definition by Johnson \&
Morgan (1953); see e.g. Massey \&
Thompson (1991). The coefficient multiplying $(B-V)$ is 
$E[U-B] / E[B-V]$ for which, as explained above, we prefer to
adopt a value of 0.80 rather than the classical 0.72.}, where the 
stars in this clump are characterized by $Q < -0.7$, roughly
corresponding to a main-sequence spectral type B3 or earlier. 

\begin{figure}
\resizebox{8.5cm}{!}{\includegraphics{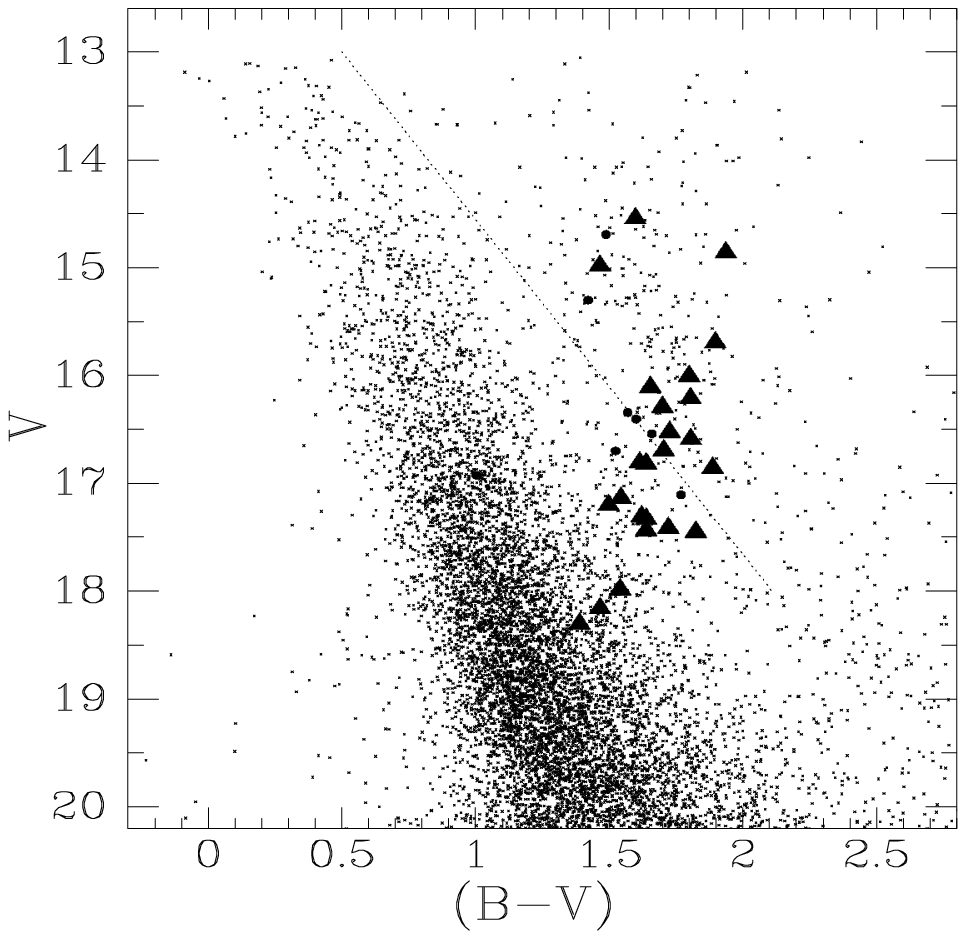}}
\par\noindent
{\bf Fig. 2.} Color-magnitude diagram showing the positions
of all the non-saturated stars in our field. The dashed line is a
reddening vector whose length corresponds to the displacement of a
star obscured by $A_V = 5$~mag. Stars with intrinsically blue
colours having $Q < -0.7$ are noted as filled circles, and the proposed WRA~751
cluster members listed in Table 1 are plotted as filled 
triangles.
\label{colmag}
\end{figure}

Figure 2 shows the position of the stars in the
$(B-V)$, $V$ diagram, with the stars having $Q < -0.7$ marked with
special symbols. The $\sigma (B-V)$, $\sigma(U-B) < 0.2$ condition 
prevents us from identifying fainter stars that may fulfill the 
$Q$-criterion, and the cutoff at faint $V$ magnitudes is thus most 
likely not real. 


\begin{table*} \caption{Likely members of the WRA~751 cluster\label{table_blue}}
\begin{tabular}{rcccccc}
\noalign{\smallskip}\hline \noalign{\smallskip}
ID & $\alpha(2000)$ & $\delta(2000)$ & $V$ & $(U-B)$ & $(B-V)$ & $Q$ \\
\noalign{\smallskip}\hline \noalign{\smallskip}
 1 & 11:08:37.6 & -60:42:46 & $14.54 \pm 0.01$ & $0.31 \pm 0.04$ & $1.59 \pm 0.00$ & $-0.96 \pm 0.04$\\
 2 & 11:08:37.8 & -60:42:49 & $16.81 \pm 0.01$ & $0.51 \pm 0.15$ & $1.63 \pm 0.02$ & $-0.79 \pm 0.15$\\
 3 & 11:08:38.0 & -60:42:35 & $17.30 \pm 0.01$ & $0.21 \pm 0.16$ & $1.62 \pm 0.03$ & $-1.08 \pm 0.16$\\
 4 & 11:08:39.9 & -60:42:15 & $18.30 \pm 0.02$ & $0.13 \pm 0.23$ & $1.38 \pm 0.05$ & $-0.97 \pm 0.24$\\
 5 & 11:08:40.1 & -60:42:18 & $14.98 \pm 0.01$ & $0.13 \pm 0.05$ & $1.46 \pm 0.01$ & $-1.03 \pm 0.05$\\
 6 & 11:08:41.2 & -60:42:34 & $16.29 \pm 0.01$ & $0.52 \pm 0.12$ & $1.69 \pm 0.01$ & $-0.83 \pm 0.12$\\
 7 & 11:08:41.6 & -60:43:18 & $16.85 \pm 0.01$ & $0.80 \pm 0.19$ & $1.88 \pm 0.02$ & $-0.70 \pm 0.19$\\
 8 & 11:08:41.6 & -60:42:41 & $14.85 \pm 0.01$ & $0.59 \pm 0.07$ & $1.93 \pm 0.01$ & $-0.95 \pm 0.07$\\
 9 & 11:08:41.7 & -60:43:01 & $18.15 \pm 0.02$ & $0.10 \pm 0.21$ & $1.46 \pm 0.04$ & $-1.07 \pm 0.22$\\
10 & 11:08:41.8 & -60:42:43 & $17.41 \pm 0.01$ & $0.64 \pm 0.22$ & $1.72 \pm 0.03$ & $-0.72 \pm 0.22$\\
11 & 11:08:42.4 & -60:42:57 & $16.52 \pm 0.01$ & $0.54 \pm 0.13$ & $1.72 \pm 0.02$ & $-0.83 \pm 0.14$\\
12 & 11:08:42.8 & -60:42:53 & $17.45 \pm 0.01$ & $0.60 \pm 0.23$ & $1.82 \pm 0.03$ & $-0.85 \pm 0.23$\\
13 & 11:08:43.1 & -60:42:50 & $16.00 \pm 0.01$ & $0.41 \pm 0.10$ & $1.79 \pm 0.02$ & $-1.02 \pm 0.11$\\
14 & 11:08:44.5 & -60:42:48 & $16.59 \pm 0.01$ & $0.61 \pm 0.15$ & $1.80 \pm 0.02$ & $-0.83 \pm 0.15$\\
15 & 11:08:45.2 & -60:42:57 & $16.80 \pm 0.01$ & $0.42 \pm 0.14$ & $1.61 \pm 0.02$ & $-0.86 \pm 0.14$\\
16 & 11:08:45.2 & -60:42:52 & $17.43 \pm 0.01$ & $0.40 \pm 0.19$ & $1.63 \pm 0.03$ & $-0.90 \pm 0.19$\\
17 & 11:08:46.0 & -60:43:00 & $17.20 \pm 0.01$ & $0.36 \pm 0.15$ & $1.49 \pm 0.02$ & $-0.83 \pm 0.16$\\
18 & 11:08:46.0 & -60:42:07 & $17.32 \pm 0.01$ & $0.54 \pm 0.19$ & $1.63 \pm 0.03$ & $-0.76 \pm 0.19$\\
19 & 11:08:46.4 & -60:41:58 & $17.13 \pm 0.01$ & $0.35 \pm 0.15$ & $1.54 \pm 0.02$ & $-0.88 \pm 0.15$\\
20 & 11:08:46.8 & -60:42:16 & $16.10 \pm 0.01$ & $0.36 \pm 0.10$ & $1.65 \pm 0.02$ & $-0.95 \pm 0.10$\\
21 & 11:08:48.9 & -60:44:23 & $15.69 \pm 0.01$ & $0.59 \pm 0.19$ & $1.89 \pm 0.05$ & $-0.92 \pm 0.28$\\
22 & 11:08:49.4 & -60:43:25 & $16.21 \pm 0.01$ & $0.47 \pm 0.22$ & $1.80 \pm 0.06$ & $-0.96 \pm 0.24$\\
23 & 11:08:50.9 & -60:42:46 & $17.98 \pm 0.01$ & $0.47 \pm 0.24$ & $1.54 \pm 0.04$ & $-0.75 \pm 0.24$\\
24 & 11:09:05.7 & -60:42:21 & $16.69 \pm 0.01$ & $0.44 \pm 0.14$ & $1.70 \pm 0.02$ & $-0.91 \pm 0.14$\\
\noalign{\smallskip}\hline
\end{tabular}
\end{table*}

\section{The WRA~751 cluster\label{cluster}}

\subsection{Imaging}

  In our initial EMMI observations we identified 20 stars fulfilling
the $Q < -0.7$ criterion. Their spatial distribution clearly revealed their
clustering around WRA~751, as 13 of them were found within a radius of only 
$76''$ from WRA~751, i.e., less than 3\% of the field imaged with EMMI.

  The deeper FORS2 observations expand this initial census up to the 24 blue 
stars in the surroundings of WRA~751 that we list in Table 1. 
Their positions are plotted in Figure 3. These stars, marked as 
triangles in Figure 2, are distributed around $B-V \simeq 1.67$, which
corresponds to an average extinction $A_V \simeq 6.1$ towards the cluster. 
Their relatively narrow range of colours indicates that the extinction 
within the cluster is rather small, probably not exceeding 
$A_{V_{\rm int}} \sim 1.5$~mag. 
Most of the blue objects listed in Table 1 and plotted in 
Figure 3 are projected within a distance of $117''$ of WRA~751, and 
the radius of the cluster is approximately 1'. Four additional stars are seen 
at larger distances, up to almost $4'$ from WRA~751. The fairly strong 
degree of concentration towards WRA~751 of the other members suggests that these
stars may not be associated to the cluster. However, both their colours and 
their magnitudes are within the range defined by the early stars closer to 
WRA~751, and we therefore consider them as possibly related to it, also given 
the rarity of stars with such photometric characteristics in the rest of the 
field covered by our observations. We have thus included them in our census 
listed in Table 1.

\begin{figure}
\resizebox{8cm}{!}{\includegraphics{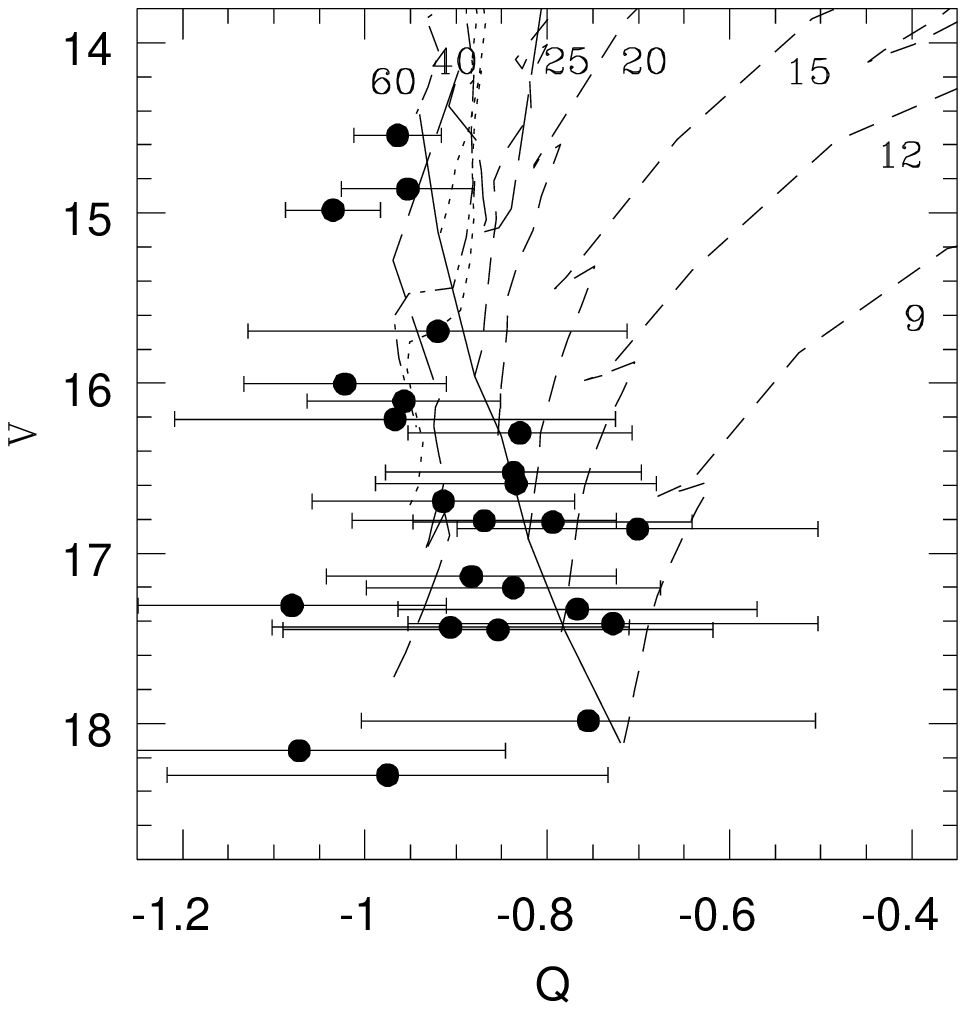}}
\par\noindent
{\bf Fig. 4.} The $V$ vs. $Q$ diagram of the intrinsically bluest stars in the 
observed fields (see text for definitions). The evolutionary tracks computed by LeJeune
\& Schaerer (2001, Case {\it e} with high mass loss and no rotation) for initial stellar 
masses of 9, 12, 15, 20 and 25 M$_{\odot}$ are shown with dashed lines. The tracks for M$_i$ = 40
and 60 M$_{\odot}$ are the dotted and dot-dashed lines, respectively. The
additional long-dashed line represents the evolutionary track for M$_i$ = 120
M$_{\odot}$. All tracks 
have been scaled to a distance of 6 kpc and reddened by A$_V$ = 6.1 (see Sect. 4.1). 
The Zero-Age Main-Sequence (ZAMS) is traced with a solid line. 
\label{QV0}
\end{figure}

  The observed $V$ vs. $Q$ diagram provides an approximation to the H-R diagram of 
the cluster, and is presented in Figure 4. Evolutionary tracks for stars of different initial
mass as given by LeJeune \& Schaerer (2001) in Case {\it e} (with high mass loss and
no rotation, see Meynet et al. 1994) are also plotted. 
Tracks corresponding to initial masses of 9, 12, 20 and 25 M$_{\odot}$ are traced with a
dashed line, while those for M$_i$ = 40 and 60 M$_{\odot}$ are shown with a dotted
and a dot-dashed line, respectively. The additional long-dashed line represents
the evolutionary track for M$_i$ = 120 M$_{\odot}$. All the tracks have been scaled to 
a distance of 6 kpc (see Sect. 5) and reddened by A$_V$ = 6.1 (see Sect. 4.1).
Their corresponding Zero-Age Main-Sequence (ZAMS) is traced with a solid line.
The large errors in the horizontal direction, mainly due to the faintness of the stars in
$U$, prevent us from identifying accurate evolutionary phases and initial
masses, and only spectroscopy can remove this degeneracy. Nevertheless, Figure 4
indicates that the cluster stars can be quite massive, spaning a range between 
9 and 120 M$_{\odot}$ (the upper limit of the evolutionary tracks), where
the lower limit of 9 M$_{\odot}$ is due to the adopted constraint of $Q <$ -0.7.

\subsection{Spectroscopy}
The spectra of the five OB candidates observed with FORS1 in July 2003 are
shown in Figure 5, where the stellar continuum has been normalized to unity.
The stars ID number is taken from Table 1.

\begin{figure}
\resizebox{9cm}{!}{\includegraphics{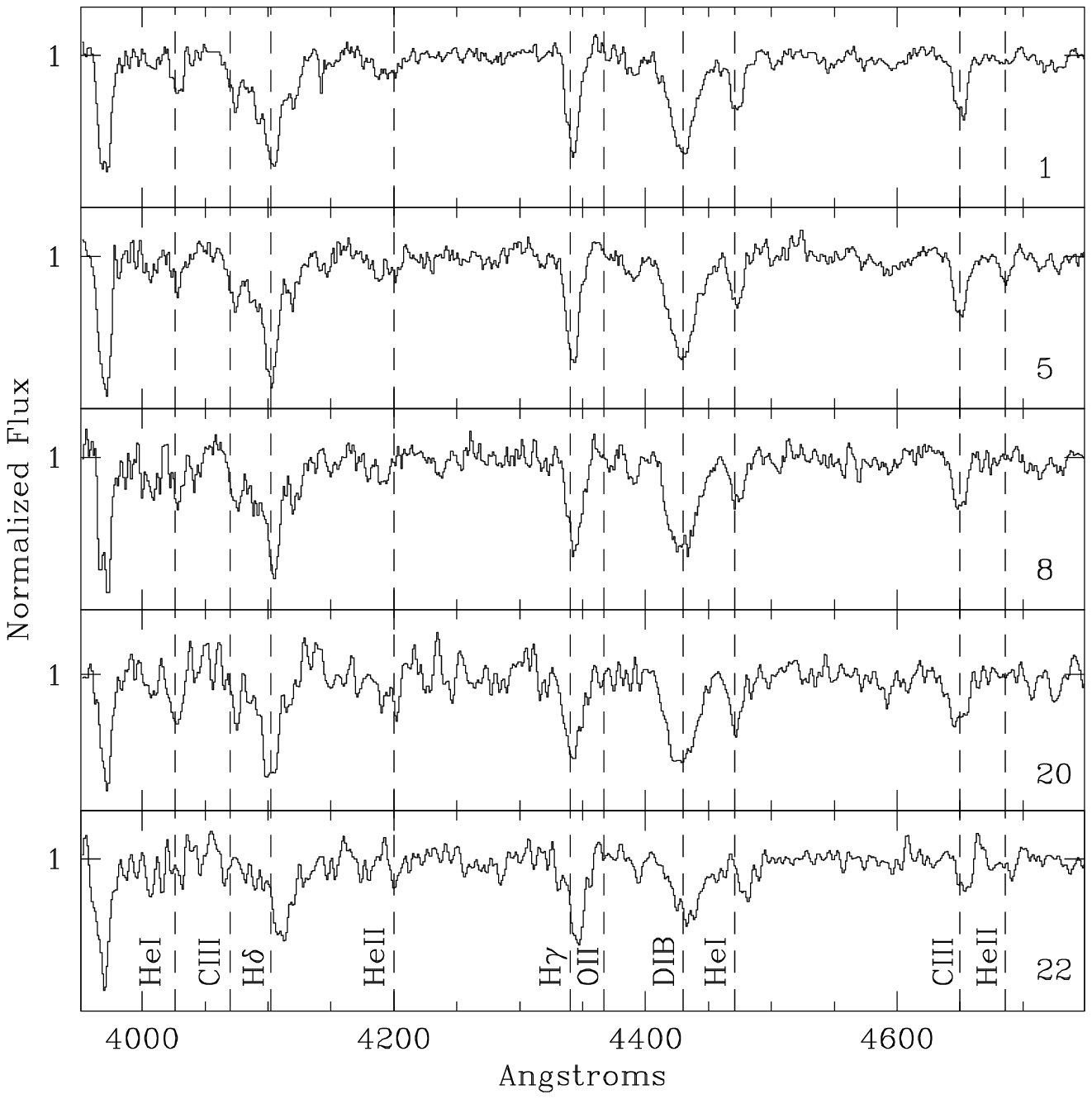}}
\par\noindent
{\bf Fig. 5.} The FORS1 spectra of five OB stars in the WRA~751 cluster. Their
stellar continuum is normalized to unity. The most relevant lines for spectral 
classification are shown together with the diffuse interstellar band in absorption (DIB)
at $\lambda \simeq$ 4430 \AA.
\end{figure}

A number of absorption lines typical of the O and B spectral types
can be easily identified: H$\gamma$, HeI $\lambda$4471, HeII $\lambda\lambda$4200,4686,
CIII $\lambda\lambda$4070,4650 and OII $\lambda$4367. The feature in
absorption at $\lambda$4428 is due to a diffuse interstellar band (DIB); its
strength does not significantly vary among the observed stars, thus
indicating quite similar reddenings. 

The spectra were classified independently by each author via comparison
with the spectral atlases of Walborn \& Fitzpatrick (1990) and Hanson (2003).
The mean spectral types derived for the targets are listed in Table 2, ranging
from as-early-as O8 to as-late-as B1. We estimate our classification to be accurate
at less than 1.5 subtypes and two luminosity classes.
We used the intrinsic $UBV$ colour terms
given by FitzGerald (1970) as a function of spectral type and luminosity class,
and derived from the observed photometry of the targets their extinction (in Table 2). 
This turns out to be in quite good agreement with the extinction estimated from 
the photometry alone (see Sect. 4.1).

\begin{table*} 
\caption{The spectral classification of five early-type stars in the
WRA~751 cluster, their temperature and absolute magnitude. A distance of 6 kpc
is assumed}
\begin{tabular}{rccccc}
\noalign{\smallskip}\hline \noalign{\smallskip}
ID & Spectral & A$_V$ & log(T$_{eff}$ [K]) & log(T$_{eff}$ [K]) & M$_V$\\
   & Type     &       & (Slesnick et al. '02) & (Massey et al. '05) &\\
\noalign{\smallskip}\hline \noalign{\smallskip}
1 & B0I & 5.67 & 4.37 & 4.47 & -5.02\\ 
5 & O9V & 5.49 & 4.60 & 4.50 & -4.40\\
8 & B1Ia & 6.57 & 4.25 & --  &-5.61\\
20 & O9I & 5.98 & 4.49 & 4.50 & -3.77\\
22 & O8V & 6.54 & 4.60 & 4.53 & -4.22\\
\noalign{\smallskip}\hline
\end{tabular}
\end{table*}

The dereddened observed $(B-V)$ colours were then used to determine 
the effective temperature ($T_{eff}$, in Table 2) 
of the targets via the empirical calibrations of Slesnick et al. (2002). 
Although these calibrations are not very sensitive to T$_{eff}$
of O and early B stars, and do not take into account the effect of
stellar wind blanketing, they extend to the B1I spectral type which is
not included in the most up-to-date temperature scales published by
Repolust et al. (2004) and Massey et al. (2005). Nevertheless, we have
reported in Table 2 the T$_{eff}$ values given by Massey et al. (2005)
for stars earlier than B0I. The difference between these two 
sets of effective temperatures is less than or comparable to the 1$\sigma$
errors on our spectral classification
(see Sect. 5). For this reason and for the sake of homogeneity, we
adopt the effective temperatures given by the empirical calibrations
of Slesnick et al. throughout Sect. 5 and in Figure 6. 

\section{Discussion}
Our observations reveal the presence of a cluster of blue stars spatially
coincident with WRA~751. Given the extreme paucity of early-type stars (with $Q <$ -0.7) 
in the observed field, we identify this cluster as the birth-site of WRA~751. A
chance alignment of a cluster of massive stars with a LBV is extremely unlikely,
given that both classes of objects are rare.

Various distance estimates can be found in literature for WRA~751. van
Genderen et al. (1992) derived a distance larger than 4 - 5 kpc from the photometry
of field stars around WRA~751. This is consistent with the more recent determination
of 6 $\pm$ 1 kpc (Nota, private communication) based
on the radial velocity of WRA~751 and its surrounding HII region. 
We have adopted the latter value in this paper; we prefer it over a distance
simply derived from the spectral types in Table 2 and the available photometry in
Table 1, since it is well known that the absolute magnitude of early-type stars
displays a large scatter within any given spectral type (Garmany 1990, Jaschek \&
G\'omez 1998). At this distance, the angular extent of the cluster translates into
a radius of 3.4 pc, with WRA~751 close to the cluster centre. Assuming that all stars
more massive than 9 M$_{\odot}$ have been detected (cf. Figure 4) and that the initial
mass function of the cluster is represented by a Miller \& Scalo's (1979) law, we
roughly estimate the cluster mass to be 2.2 $\times$ 10$^3$ M$_{\odot}$, i.e. 
about one order of magnitude larger than the Trapezium cluster in Orion (Herbig \&
Terndrup 1986).

The H-R diagram of the observed five OB stars in the WRA~751 cluster is plotted in
Figure 6. The 1$\sigma$ errors on $T_{eff}$ simply reflect the 
uncertainty of our spectral classification, while those on M$_V$ take
also into account the uncertainty on the adopted distance, the photometric errors
and the uncertainty in $R_V$ (see Sect. 3). The evolutionary tracks
(dashed lines) are taken from LeJeune \& Schaerer (2001, Case {\it e} with high
mass loss and no rotation, see also Meynet et al. 1994); their ZAMS is represented with a
solid line. We have added to the observed OB stars WRA~751 itself, for which we have
assumed a mean, observed magnitude $V$ = 12.14, an extinction $A_V$ = 5.58 mag and an O9.5I
spectral type (cf. Hu et al. 1990). The effective temperature of WRA~751 has been
derived once again from the empirical calibrations of Slesnick et al. 

The position of the targets in the H-R diagram indicates that the observed OB stars span
a mass range between 12 and 40 M$_{\odot}$, although the uncertainty on our spectral
classification allows masses larger than 40 M$_{\odot}$ in the case of the two O8V and
O9V stars ($\#$ 22 and 5 respectively). The progenitor of WRA~751 may have
been as massive as $\sim$ 80 M$_{\odot}$. Since the brighter and bluer stars were selected for the
spectroscopic follow-up, we are confident that no star earlier than O8 is present in the
cluster. If the absence of stars earlier than O8 were due to
their evolution off the main sequence, this would set a lower limit on
the age of the cluster of about 4 Myr, which turns out to be
consistent with the current evolutionary phase of WRA~751. We note
however that the absence of such early-type stars may also be simply due
to an insufficient sampling of the upper end of the main sequence (i.e.,
small-number statistics) given the limited mass that we estimate for the
cluster. Thus, a younger age cannot be ruled out, although the
classification of three late O/early B stars with luminosity class I
argues for significant evolution implying an age of a few million
years.

The data presented here offer a first glimpse of the nature of the birth-cluster of WRA~751,
and need to be complemented with additional  observations,
leading to an accurate determination of its distance, stellar content and age.
Its identification and study are an important step towards our
understanding of the evolution of WRA~751 into the Luminous Blue Variable phase.

\begin{figure}
\resizebox{9.cm}{!}{\includegraphics{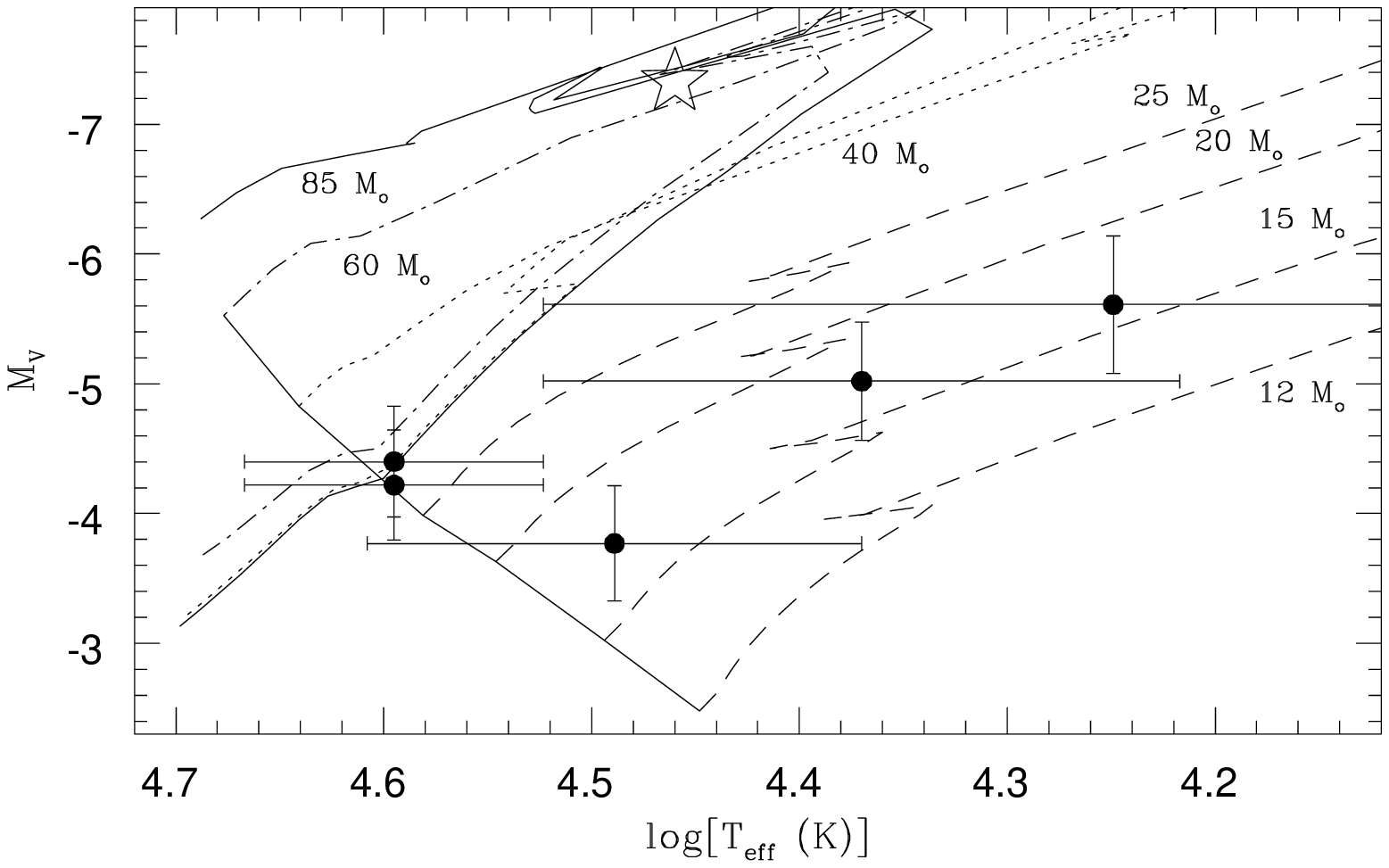}}
\par\noindent
{\bf Fig. 6.} The H-R diagram of the newly discovered OB stars around WRA~751, 
at an assumed distance of 6 kpc (filled circles). The LBV star WRA~751 is shown with 
an open star symbol. Evolutionary tracks are from LeJeune \& Schaerer (2001, Case {\it e}
with high mass loss and no rotation).
Tracks corresponding to initial masses of 12, 15, 20 and 25 M$_{\odot}$ are traced
as dashed lines, while those referring to M$_i$ = 40, 60 and 85 M$_{\odot}$ are
shown with a dotted, dot-dashed and solid line, respectively.
\end{figure}

\begin{acknowledgements}
We acknowledge an anonymous referee whose comments and suggestions
improved the paper.
We are pleased to thank the staff of the La Silla Observatory
for their support during our NTT observations, especially Mr.
Duncan Castex and Ms. M\'onica Castillo. We also thank the ESO
User Support Department for their valuable assistance in the
preparation of our spectroscopic observations, the ESO Data Flow 
Operations Group for their delivery of a data package of 
excellent quality, and the Paranal Science Operations staff in 
charge of the execution of the Service Mode observations. The 
allocation of time for the follow-up spectroscopy by the ESO 
Director General Discretionary Time Committee is gratefully 
acknowledged.

\end{acknowledgements}

\clearpage
\begin{figure*}
\resizebox{14cm}{!}{\includegraphics{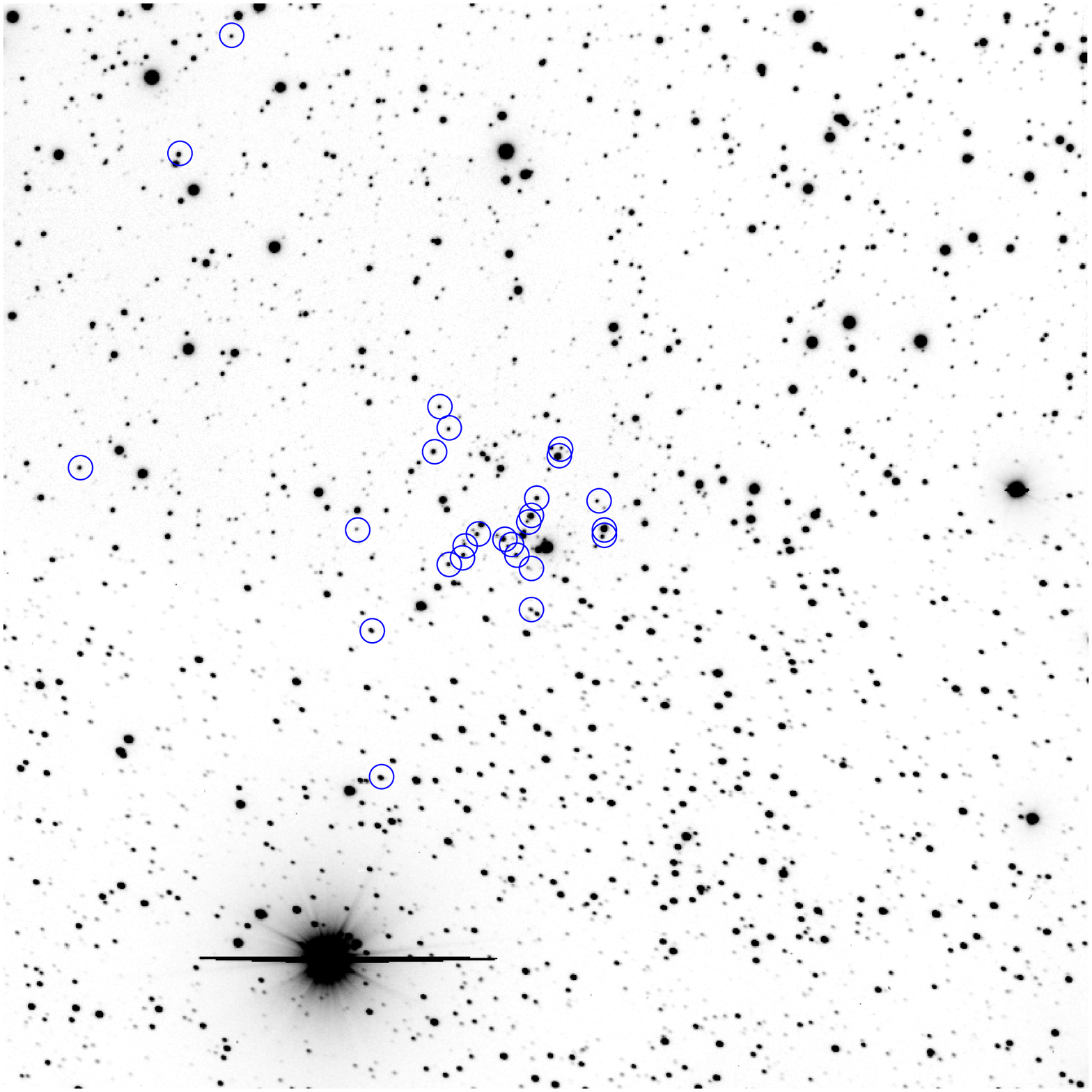}}
\par\noindent
{\bf Fig. 3.} A $B$-band image of the field around WRA~751, with the stars
having $Q < -0.7$ marked (see Sect. 3). The field
covers $7'.06 \times 7'.06$ and WRA~751 is at its center.  North is up and
East to the left.
\label{field}
\end{figure*}

\end{document}